%% file: main.tex
\pdfoutput=1
\documentclass[12pt,a4paper]{article}
\usepackage{ifthen} % for conditional statements
\newboolean{pdflatex}
\setboolean{pdflatex}{true} % False for eps figures 
\usepackage{footmisc}
\newboolean{articletitles}
\setboolean{articletitles}{true} % False removes titles in references

\newboolean{uprightparticles}
\setboolean{uprightparticles}{false} %True for upright particle symbols

\newboolean{inbibliography}
\setboolean{inbibliography}{false} %True once you enter the bibliography

\input{preamble}
\usepackage{longtable} % only for template; not usually to be used in PAPERs
\usepackage{mathtools}

\def\JpsipK {\decay{\Lb}{ \jpsi\proton\Km}}
\def\jpsipK {\JpsipK}
\def\jpsipk {\JpsipK}
\def\pkjpsi {\JpsipK}
\def\Jpsippi {\decay{\Lb}{ \jpsi\proton\pim}}
\def\jpsippi {\Jpsippi}
\def\ppijpsi {\Jpsippi}
\def\pKmumu {\decay{\Lb}{ \proton\Km\mumu}}
\def\pkmumu {\pKmumu}
\def\ppimumu {\decay{\Lb}{ \proton\pim\mumu}}

\begin{document}

%%%%%%%%%%%%%%%%%%%%%%%%%
%%%%% Title     %%%%%%%%%
%%%%%%%%%%%%%%%%%%%%%%%%%
\renewcommand{\thefootnote}{\fnsymbol{footnote}}
\setcounter{footnote}{1}

% %%%%%%% CHOOSE TITLE PAGE--------
%\onecolumn
%\input{title-LHCb-INT}
%\input{title-LHCb-ANA}
%\input{title-LHCb-CONF}
\input{title-LHCb-PAPER}

%\twocolumn
% %%%%%%%%%%%%% ---------

\renewcommand{\thefootnote}{\arabic{footnote}}
\setcounter{footnote}{0}

%%%%%%%%%%%%%%%%%%%%%%%%%%%%%%%%
%%%%%  Table of Content   %%%%%%
%%%%%%%%%%%%%%%%%%%%%%%%%%%%%%%%
%%%% Uncomment next 2 lines if desired
%\tableofcontents
%\cleardoublepage

%%%%%%%%%%%%%%%%%%%%%%%%%
%%%%% Main text %%%%%%%%%
%%%%%%%%%%%%%%%%%%%%%%%%%

\pagestyle{plain} % restore page numbers for the main text
\setcounter{page}{1}
\pagenumbering{arabic}

%% Uncomment during review phase. 
%% Comment before a final submission.
% \linenumbers

% You can include short sections directly in the main tex file.
% However, for larger papers it is desirable to split the text into
% several semiautonomous files, which can be revised independently.
% This is especially useful when developing a document in
% collaboration with several people, since then different parts can be
% edited independently.  This type of file organization is shown here.
% 
\renewcommand{\thefootnote}{\fnsymbol{footnote}}

\input{introduction}

\input{detector}

\input{selection}

\input{normalisation}

\input{results}

\input{systematics}

\input{conclusions}
%\input{supplementary-app}
% Do not include this in analysis note and conference reports
\input{acknowledgements}

%\input{appendix}

% This should be taken out in the final paper
% \input{supplementary-app}

\addcontentsline{toc}{section}{References}
\ifx\mcitethebibliography\mciteundefinedmacro
\PackageError{LHCb.bst}{mciteplus.sty has not been loaded}
{This bibstyle requires the use of the mciteplus package.}\fi
\providecommand{\href}[2]{#2}

%% \setboolean{inbibliography}{true}
%% \bibliographystyle{LHCb}
%% \bibliography{main,LHCb-PAPER,LHCb-CONF,LHCb-DP,LHCb-TDR}

\newpage

% Author List ----------------------------                                                                                                                                                                                                                                                                                                
%  You need to get a new author list!                                                                                                                                                                                                                                                                                                    

%\input{LHCb_HD_authorlist_2014-06-20}
 
\newpage
\input{LHCb_Authorship_flat_27-Sep-2016.tex}

\end{document}

%% file: preamble.tex
\usepackage[top=1in, bottom=1.25in, left=1in, right=1in]{geometry}

\usepackage{microtype}
\usepackage{lineno}  % for line numbering during review
\usepackage{xspace} % To avoid problems with missing or double spaces after
\usepackage{caption} %these three command get the figure and table captions automatically small

\usepackage{graphicx}  % to include figures (can also use other packages)
\usepackage{color}
\usepackage{colortbl}
\graphicspath{{./figs/}} % Make Latex search fig subdir for figures

\usepackage{amsmath} % Adds a large collection of math symbols
\usepackage{amssymb}
\usepackage{amsfonts}
\usepackage{upgreek} % Adds in support for greek letters in roman typeset

\newcommand*\patchAmsMathEnvironmentForLineno[1]{%
\expandafter\let\csname old#1\expandafter\endcsname\csname #1\endcsname
\expandafter\let\csname oldend#1\expandafter\endcsname\csname
end#1\endcsname
 \renewenvironment{#1}%
   {\linenomath\csname old#1\endcsname}%
   {\csname oldend#1\endcsname\endlinenomath}%
}
\newcommand*\patchBothAmsMathEnvironmentsForLineno[1]{%
  \patchAmsMathEnvironmentForLineno{#1}%
  \patchAmsMathEnvironmentForLineno{#1*}%
}
\AtBeginDocument{%
\patchBothAmsMathEnvironmentsForLineno{equation}%
\patchBothAmsMathEnvironmentsForLineno{align}%
\patchBothAmsMathEnvironmentsForLineno{flalign}%
\patchBothAmsMathEnvironmentsForLineno{alignat}%
\patchBothAmsMathEnvironmentsForLineno{gather}%
\patchBothAmsMathEnvironmentsForLineno{multline}%
\patchBothAmsMathEnvironmentsForLineno{eqnarray}%
}

\usepackage{hyperref}    % Hyperlinks in references
\usepackage[all]{hypcap} % Internal hyperlinks to floats.

\input{lhcb-symbols-def} % Add in the predefined LHCb symbols

\usepackage{cite} % Allows for ranges in citations
\usepackage{mciteplus}

%% file: lhcb-symbols-def.tex
\usepackage{xspace} 
\usepackage{upgreek}

\def\lhcb {\mbox{LHCb}\xspace}

\def\MagUp {\mbox{\em Mag\kern -0.05em Up}\xspace}

\ifthenelse{\boolean{uprightparticles}}%
{

 \def\Pmu         {\ensuremath{\upmu}\xspace}

 \def\Ppi         {\ensuremath{\uppi}\xspace}

 \def\Ppsi        {\ensuremath{\uppsi}\xspace}

 \def\PDelta      {\ensuremath{\Delta}\xspace}                 
 \def\PXi      {\ensuremath{\Xi}\xspace}                 
 \def\PLambda      {\ensuremath{\Lambda}\xspace}                 
 \def\PSigma      {\ensuremath{\Sigma}\xspace}                 
 \def\POmega      {\ensuremath{\Omega}\xspace}                 
 \def\PUpsilon      {\ensuremath{\Upsilon}\xspace}                 
 
 \def\PB      {\ensuremath{\mathrm{B}}\xspace}                 
                  
 \def\PD      {\ensuremath{\mathrm{D}}\xspace}

 \def\PJ      {\ensuremath{\mathrm{J}}\xspace}                 
 \def\PK      {\ensuremath{\mathrm{K}}\xspace}

 \def\Pb      {\ensuremath{\mathrm{b}}\xspace}                 
 \def\Pc      {\ensuremath{\mathrm{c}}\xspace}                 
 \def\Pd      {\ensuremath{\mathrm{d}}\xspace}

 \def\Pi      {\ensuremath{\mathrm{i}}\xspace}

 \def\Pp      {\ensuremath{\mathrm{p}}\xspace}

 \def\Ps      {\ensuremath{\mathrm{s}}\xspace}                 
 \def\Pt      {\ensuremath{\mathrm{t}}\xspace}

}
{

 \def\Pmu         {\ensuremath{\mu}\xspace}

 \def\Ppi         {\ensuremath{\pi}\xspace}

 \def\Ppsi        {\ensuremath{\psi}\xspace}                 
                  
 \mathchardef\PDelta="7101
 \mathchardef\PXi="7104
 \mathchardef\PLambda="7103
 \mathchardef\PSigma="7106
 \mathchardef\POmega="710A
 \mathchardef\PUpsilon="7107
                  
 \def\PB      {\ensuremath{B}\xspace}                 
                  
 \def\PD      {\ensuremath{D}\xspace}

 \def\PJ      {\ensuremath{J}\xspace}                 
 \def\PK      {\ensuremath{K}\xspace}

 \def\Pb      {\ensuremath{b}\xspace}                 
 \def\Pc      {\ensuremath{c}\xspace}                 
 \def\Pd      {\ensuremath{d}\xspace}

 \def\Pi      {\ensuremath{i}\xspace}

 \def\Pp      {\ensuremath{p}\xspace}

 \def\Ps      {\ensuremath{s}\xspace}                 
 \def\Pt      {\ensuremath{t}\xspace}

}

\makeatletter
\ifcase \@ptsize \relax% 10pt
  \newcommand{\miniscule}{\@setfontsize\miniscule{4}{5}}% \tiny: 5/6
\or% 11pt
  \newcommand{\miniscule}{\@setfontsize\miniscule{5}{6}}% \tiny: 6/7
\or% 12pt
  \newcommand{\miniscule}{\@setfontsize\miniscule{5}{6}}% \tiny: 6/7
\fi
\makeatother

\DeclareRobustCommand{\optbar}[1]{\shortstack{{\miniscule (\rule[.5ex]{1.25em}{.18mm})}
  \\ [-.7ex] $#1$}}

\def\mup        {{\ensuremath{\Pmu^+}}\xspace}
\def\mun        {{\ensuremath{\Pmu^-}}\xspace} % muon negative (\mum is taken)
\def\mumu       {{\ensuremath{\Pmu^+\Pmu^-}}\xspace}

\def\dquark    {{\ensuremath{\Pd}}\xspace}

\def\squark    {{\ensuremath{\Ps}}\xspace}

\def\cquark    {{\ensuremath{\Pc}}\xspace}

\def\bquark    {{\ensuremath{\Pb}}\xspace}

\def\tquark    {{\ensuremath{\Pt}}\xspace}

\def\pion   {{\ensuremath{\Ppi}}\xspace}

\def\pim    {{\ensuremath{\pion^-}}\xspace}

\def\kaon    {{\ensuremath{\PK}}\xspace}
  \def\Kbar    {{\kern 0.2em\overline{\kern -0.2em \PK}{}}\xspace}

\def\KorKbar    {\kern 0.18em\optbar{\kern -0.18em K}{}\xspace}

\def\Km      {{\ensuremath{\kaon^-}}\xspace}

  \def\Dbar    {{\kern 0.2em\overline{\kern -0.2em \PD}{}}\xspace}

\def\DorDbar    {\kern 0.18em\optbar{\kern -0.18em D}{}\xspace}

\def\B       {{\ensuremath{\PB}}\xspace}
\def\Bbar    {{\ensuremath{\kern 0.18em\overline{\kern -0.18em \PB}{}}}\xspace}

\def\BorBbar    {\kern 0.18em\optbar{\kern -0.18em B}{}\xspace}
\def\Bz      {{\ensuremath{\B^0}}\xspace}
\def\Bzb     {{\ensuremath{\Bbar{}^0}}\xspace}

\def\Bs      {{\ensuremath{\B^0_\squark}}\xspace}

\def\jpsi     {{\ensuremath{{\PJ\mskip -3mu/\mskip -2mu\Ppsi\mskip 2mu}}}}
\def\psitwos  {{\ensuremath{\Ppsi{(2S)}}}\xspace}

  \def\Y#1S{\ensuremath{\PUpsilon{(#1S)}}\xspace}% no space before {...}!

\def\proton      {{\ensuremath{\Pp}}\xspace}

\def\Lz          {{\ensuremath{\PLambda}}\xspace}
\def\Lbar        {{\ensuremath{\kern 0.1em\overline{\kern -0.1em\PLambda}}}\xspace}
\def\LorLbar    {\kern 0.18em\optbar{\kern -0.18em \PLambda}{}\xspace}

\def\Lb      {{\ensuremath{\Lz^0_\bquark}}\xspace}

\def\Lc      {{\ensuremath{\Lz^+_\cquark}}\xspace}

\newcommand{\decay}[2]{\ensuremath{#1\!\to #2}\xspace}         % {\Pa}{\Pb \Pc}

\def\to                 {\ensuremath{\rightarrow}\xspace}

\def\qsq       {{\ensuremath{q^2}}\xspace}

\def\Vtd  {{\ensuremath{V_{\tquark\dquark}}}\xspace}

\def\Vts  {{\ensuremath{V_{\tquark\squark}}}\xspace}

\def\AT#1     {\ensuremath{A_{\mathrm{T}}^{#1}}\xspace}           % 2

\def\C#1      {\ensuremath{\mathcal{C}_{#1}}\xspace}                       % 9
\def\Cp#1     {\ensuremath{\mathcal{C}_{#1}^{'}}\xspace}                    % 7
\def\Ceff#1   {\ensuremath{\mathcal{C}_{#1}^{\mathrm{(eff)}}}\xspace}        % 9  
\def\Cpeff#1  {\ensuremath{\mathcal{C}_{#1}^{'\mathrm{(eff)}}}\xspace}       % 7
\def\Ope#1    {\ensuremath{\mathcal{O}_{#1}}\xspace}                       % 2
\def\Opep#1   {\ensuremath{\mathcal{O}_{#1}^{'}}\xspace}                    % 7

\newcommand{\tev}{\ifthenelse{\boolean{inbibliography}}{\ensuremath{~T\kern -0.05em eV}\xspace}{\ensuremath{\mathrm{\,Te\kern -0.1em V}}}\xspace}
\newcommand{\gev}{\ensuremath{\mathrm{\,Ge\kern -0.1em V}}\xspace}
\newcommand{\mev}{\ensuremath{\mathrm{\,Me\kern -0.1em V}}\xspace}
\newcommand{\kev}{\ensuremath{\mathrm{\,ke\kern -0.1em V}}\xspace}
\newcommand{\ev}{\ensuremath{\mathrm{\,e\kern -0.1em V}}\xspace}
\newcommand{\gevc}{\ensuremath{{\mathrm{\,Ge\kern -0.1em V\!/}c}}\xspace}
\newcommand{\mevc}{\ensuremath{{\mathrm{\,Me\kern -0.1em V\!/}c}}\xspace}
\newcommand{\gevcc}{\ensuremath{{\mathrm{\,Ge\kern -0.1em V\!/}c^2}}\xspace}
\newcommand{\gevgevcccc}{\ensuremath{{\mathrm{\,Ge\kern -0.1em V^2\!/}c^4}}\xspace}
\newcommand{\mevcc}{\ensuremath{{\mathrm{\,Me\kern -0.1em V\!/}c^2}}\xspace}

\def\mum  {\ensuremath{{\,\upmu\mathrm{m}}}\xspace}

\def\invfb   {\ensuremath{\mbox{\,fb}^{-1}}\xspace}

\def\gsim{{~\raise.15em\hbox{$>$}\kern-.85em
          \lower.35em\hbox{$\sim$}~}\xspace}
\def\lsim{{~\raise.15em\hbox{$<$}\kern-.85em
          \lower.35em\hbox{$\sim$}~}\xspace}

\def\sPlot{\mbox{\em sPlot}\xspace}
\def\ptot       {\mbox{$p$}\xspace}
\def\pt         {\mbox{$p_{\mathrm{ T}}$}\xspace}

\def\evtgen     {\mbox{\textsc{EvtGen}}\xspace}

\def\geant      {\mbox{\textsc{Geant4}}\xspace}

\def\photos     {\mbox{\textsc{Photos}}\xspace}

\def\pythia     {\mbox{\textsc{Pythia}}\xspace}

\def\tell1  {TELL1\xspace}
\def\ukl1   {UKL1\xspace}

 \def\tw{0.487\pm0.022}

 \def\twq2{0.528\pm0.042}
  
  \def\Nyield{1017\pm41}
            \def\Syield{22\pm6}
                     \def\SIG{5.5}
                                 
                                       \def\BFV{(6.9 \pm 1.9 \pm 1.1^{+1.3}_{-1.0})\times 10^{-8}}
                                       

%% file: title-LHCb-PAPER.tex
% $Id: title-LHCb-PAPER.tex 103106 2017-01-26 21:16:34Z patel $
% ===============================================================================
% Purpose: LHCb-PAPER journal paper title page template
% Author: 
% Created on: 2010-09-25
% ===============================================================================

%%%%%%%%%%%%%%%%%%%%%%%%%
%%%%%  TITLE PAGE  %%%%%%
%%%%%%%%%%%%%%%%%%%%%%%%%
\begin{titlepage}
\pagenumbering{roman}

% Header ---------------------------------------------------
\vspace*{-1.5cm}
\centerline{\large EUROPEAN ORGANIZATION FOR NUCLEAR RESEARCH (CERN)}
\vspace*{1.5cm}
\noindent
\begin{tabular*}{\linewidth}{lc@{\extracolsep{\fill}}r@{\extracolsep{0pt}}}
\ifthenelse{\boolean{pdflatex}}% Logo format choice
{\vspace*{-2.7cm}\mbox{\!\!\!\includegraphics[width=.14\textwidth]{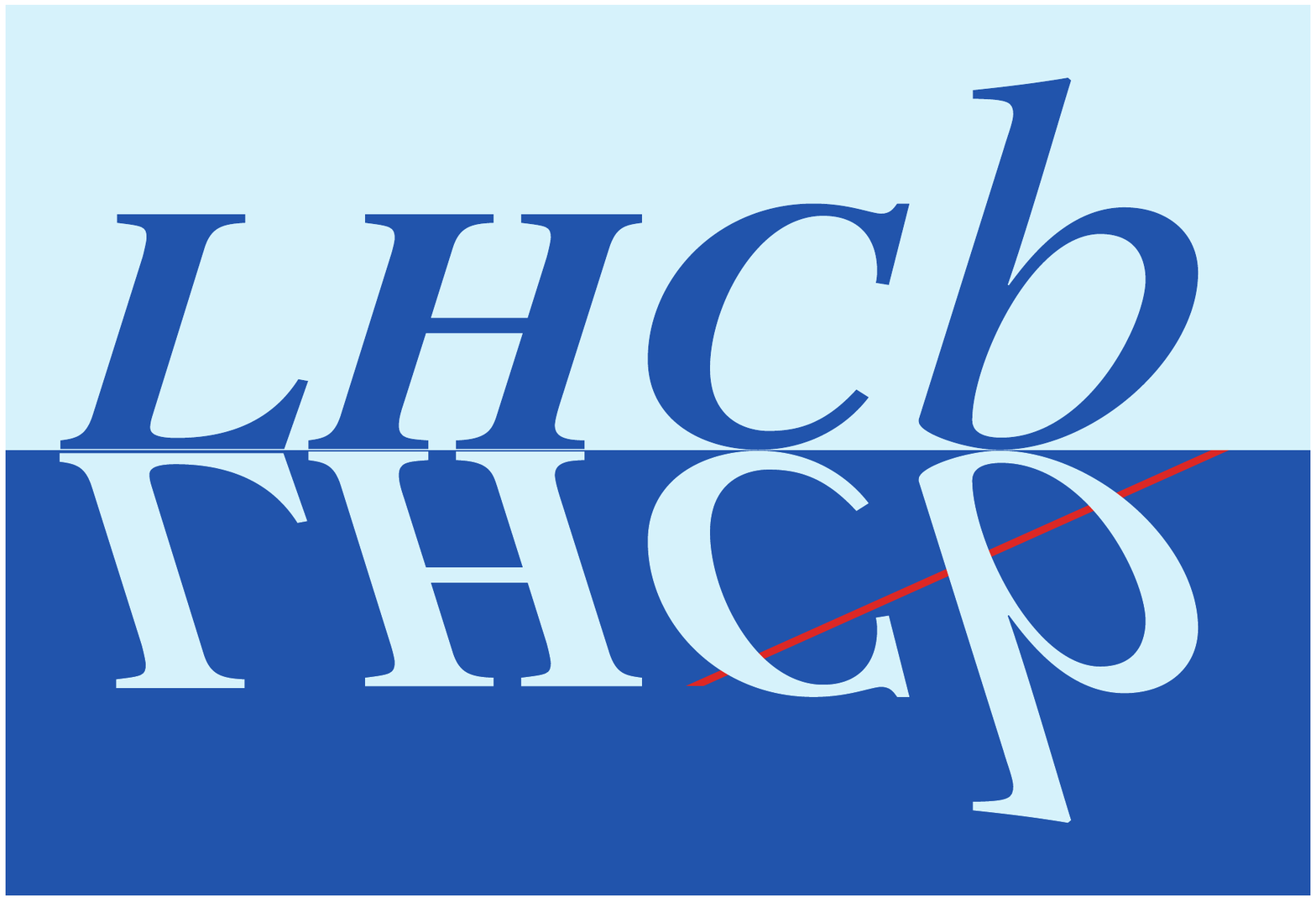}} & &}%
{\vspace*{-1.2cm}\mbox{\!\!\!\includegraphics[width=.12\textwidth]{lhcb-logo.eps}} & &}%
\\
 & & CERN-EP-2016-312 \\  % ID 
 & & LHCb-PAPER-2016-049 \\  % ID 
 & & 30 January 2017 \\ % Date - Can also hardwire e.g.: 23 March 2010
 & & \\
% not in paper \hline
\end{tabular*}

\vspace*{4.0cm}

% Title --------------------------------------------------
{\normalfont\bfseries\boldmath\huge
\begin{center}
  Observation of the suppressed decay \ppimumu
\end{center}
}

\vspace*{2.0cm}

% Authors -------------------------------------------------
\begin{center}
%In the footnote, replace 'paper' by 'Letter' in case of submission to PRL or PLB 
The LHCb collaboration\footnote{Authors are listed at the end of this paper.}
\end{center}

\vspace{\fill}

% Abstract -----------------------------------------------
\begin{abstract}
  \noindent  
The suppressed decay \ppimumu, excluding the \jpsi~and \mbox{\psitwos\to\mumu} resonances, is observed for the first time with a significance of $\SIG$ standard deviations.
The analysis is performed with proton-proton collision data corresponding to an integrated luminosity of 3\invfb collected with the LHCb experiment. The \mbox{\ppimumu} branching fraction is measured relative to the \mbox{\Lb\to\jpsi(\to\mumu)\proton{}\pim} branching fraction giving
\begin{align}
\nonumber
   \frac{\mathcal{B}(\ppimumu)}{\mathcal{B}({\decay{\Lb}{ \jpsi(\to\mumu)\proton\pim}})} &=  0.044\pm0.012\pm0.007,
\end{align}
where the first uncertainty is statistical and the second is systematic. 
This is the first observation of a $\bquark\to\dquark$ transition in a baryonic decay.

\end{abstract}

\vspace*{2.0cm}

\begin{center}
  Published in JHEP 04 (2017) 029
\end{center}

\vspace{\fill}

{\footnotesize 
\centerline{\copyright~CERN on behalf of the \lhcb collaboration, licence \href{http://creativecommons.org/licenses/by/4.0/}{CC-BY-4.0}.}}
\vspace*{2mm}

\end{titlepage}

%%%%%%%%%%%%%%%%%%%%%%%%%%%%%%%%
%%%%%  EOD OF TITLE PAGE  %%%%%%
%%%%%%%%%%%%%%%%%%%%%%%%%%%%%%%%

%  empty page follows the title page ----
\newpage
\setcounter{page}{2}
\mbox{~}
%\newpage
%
%% Author List ----------------------------
%%  You need to get a new author list!
%\input{LHCb_authorlist.tex}
%
%The author list for journal publications is provided by the Membership Committee shortly after 'approval to go to paper' has been given.
%%It will be made available on the page
%%\verb!http://www.physik.uzh.ch/~strauman/forMemCo/LHCb-PAPER-XXXX-XXX/! .
%It will be sent to you by email shortly after a paper number has beens assigned.
%The author list should be included already at first circulation, 
%to allow new members of the collaboration to verify whether they have been included correctly.
%Occasionally a misspelled name is corrected or associated institutions become full members.
%In that case, a new author list will be sent to you.
%In case line numbering doesn't work well after including the authorlist, try moving the \verb!\bigskip! after the last author to a separate line.
%
%
%The authorship for Conference Reports should be ``The LHCb
%  collaboration'', with a footnote giving the name(s) of the contact
%  author(s), but without the full list of collaboration names.

\cleardoublepage

%% file: introduction.tex
% $Id: introduction.tex 103182 2017-01-27 20:57:03Z mitesh $

\section{Introduction}
\label{sec:Introduction}

\begin{figure}[!htb]
  \centering
  \includegraphics[width=0.7\linewidth, clip = true, trim = 0 16cm 0 0]{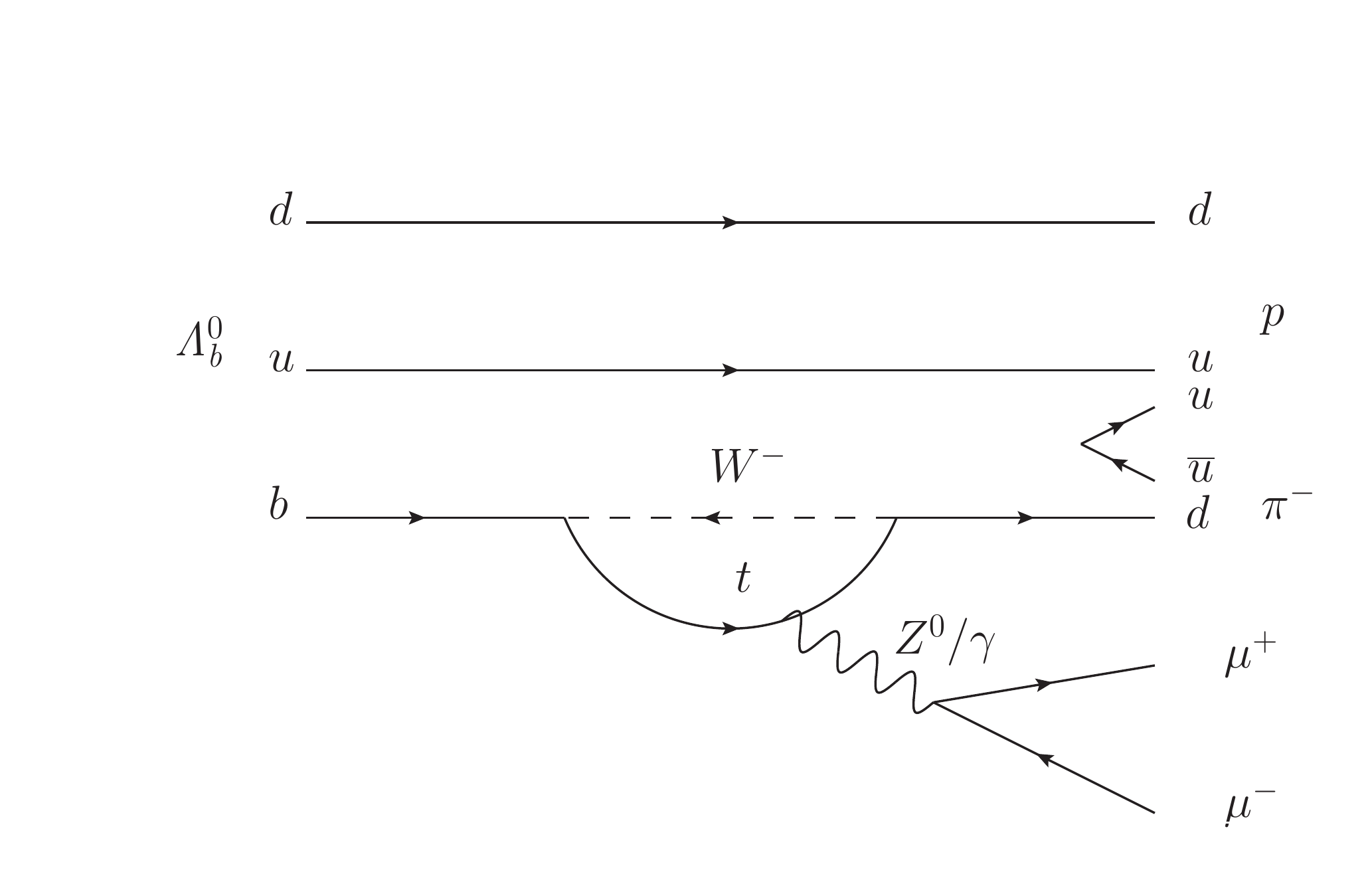}
  \caption{One of the lowest-order diagrams for the decay \ppimumu.}
  \label{fig:DecayDiagram}
\end{figure}

The decay of the \Lb baryon into the $p\pi^{-}\mumu$ final state, where the muons do not originate from a hadronic resonance, is mediated by a $\bquark\to\dquark$ transition. 
Such decays are highly suppressed in the Standard Model (SM), as the leading order amplitudes 
are described by loop diagrams and are 
also suppressed by the relevant Cabibbo-Kobayshi-Maskawa (CKM) factors. This suppression is not necessarily present in extensions to the SM, and such decays are therefore sensitive to contributions from new particles. 
One of the lowest-order diagrams for the decay \ppimumu is shown in Fig.~\ref{fig:DecayDiagram}. 

The branching fraction
of the decay\footnote[3]{The inclusion of charge-conjugate processes is implied throughout this paper.} \ppimumu is expected to be of $\mathcal{O}(10^{-8})$.
Together with the relevant form factors, a measurement of this branching fraction
 with respect to that of the analogous $\bquark\to\squark$ transition,
\pKmumu, would allow the ratio of CKM elements $|V_{td}|/|V_{ts}|$ to be determined. 
Comparing the value of $|V_{td}|/|V_{ts}|$  from these  processes with that measured via mixing processes would test the Minimal Flavour Violation hypothesis \cite{MFV,MFV2, LHCb-PAPER-2015-035}.

At present, no form-factor calculations have been made for the \ppimumu and \pKmumu channels due to the complicated hadronic structure in the proton-meson systems.
However, recent advances in lattice calculations~\cite{multihadronlattice} 
could make this possible in the future. 

This paper describes a search for the decay \ppimumu, using proton-proton collision data corresponding to an integrated luminosity of 3\invfb. The data were collected with the LHCb experiment at centre-of-mass energies of 7 and 8\tev. The branching fraction is 
determined relative to that of the tree-level decay, \Lb\to\jpsi(\to\mumu)\proton\pim, denoted as \jpsippi hereafter, which has been measured with a  precision of 15\%~\cite{LHCB-PAPER-2015-032, LHCB-PAPER-2014-020}.

%% file: detector.tex
\section{Detector and simulation}
\label{sec:Detector}

The \lhcb detector~\cite{Alves:2008zz,LHCb-DP-2014-002} is a single-arm forward
spectrometer covering the \mbox{pseudorapidity} range $2<\eta <5$,
designed for the study of particles containing \bquark or \cquark
quarks. The detector includes a high-precision tracking system
consisting of a silicon-strip vertex detector surrounding the $pp$
interaction region, a large-area silicon-strip detector located
upstream of a dipole magnet with a bending power of about
$4{\mathrm{\,Tm}}$, and three stations of silicon-strip detectors and straw
drift tubes placed downstream of the magnet.
The tracking system provides a measurement of momentum, \ptot, of charged particles with
a relative uncertainty that varies from 0.5\% at low momentum to 1.0\% at 200\gevc.
The minimum distance of a track to a primary vertex (PV), the impact parameter, 
is measured with a resolution of $(15+29/\pt)\mum$,
where \pt is the component of the momentum transverse to the beam, in\,\gevc.
Different types of charged hadrons are distinguished using information
from two Ring-Imaging Cherenkov (RICH) detectors. 
Photons, electrons and hadrons are identified by a calorimeter system consisting of
scintillating-pad and preshower detectors, an electromagnetic
calorimeter and a hadronic calorimeter. Muons are identified by a
system composed of alternating layers of iron and multiwire
proportional chambers. The online event selection is performed by a trigger~\cite{LHCb-DP-2012-004}, 
which consists of a hardware stage, based on information from the calorimeter and muon
systems, followed by a software stage, which applies a full event
reconstruction.

Simulated events are used to optimise selection criteria and calculate 
the relative efficiency between the signal and normalisation channels.
In the simulation, $pp$ collisions are generated using
\pythia~\cite{Sjostrand:2006za,*Sjostrand:2007gs} 
 with a specific \lhcb
configuration~\cite{LHCb-PROC-2010-056}.  Decays of hadronic particles
are described by \evtgen~\cite{Lange:2001uf}, in which final-state
radiation is generated using \photos~\cite{Golonka:2005pn}. The
interaction of the generated particles with the detector, and its response,
are implemented using the \geant
toolkit~\cite{Allison:2006ve, *Agostinelli:2002hh}, as described in Ref.~\cite{LHCb-PROC-2011-006}.

%% file: selection.tex
% $Id: introduction.tex 87303 2016-02-08 13:44:29Z lafferty $

\section{Selection}
\label{sec:Selection}

The \ppimumu signal candidates are first required
      to pass the hardware trigger, which selects events containing at least
      one muon with \pt greater than 1.48\gevc in the 7\tev data or
      $\pt>1.76\gevc$ in the 8\tev data.  In the subsequent software
      trigger, at least one of the final-state particles is required to 
      have $\pt>1.7\gevc$ in the 7\tev data or $\pt>1.6\gevc$ in the 8\tev 
      data. For muon candidates, a softer requirement of 
      $\pt>1.0\gevc$ is applied. The final-state particles that 
      satisfy these transverse momentum criteria are also required 
      to have an impact parameter larger than $100\mum$ with respect 
      to all PVs in the event. Finally, the tracks of two or more of 
      the final-state particles are required to form a vertex that is 
      significantly displaced from all PVs.
      
Signal candidates are reconstructed by combining two oppositely-charged 
muons with two additional tracks that are identified as a proton 
and a pion using particle identification (PID) information that comes primarily from the RICH detectors. All final-state particles are required to
have a good-quality track fit and to be inconsistent with originating from 
a PV. The pion (proton) candidates are required to have $\pt > 0.4\gevc$ 
and momentum greater than 2.0 (7.5)\gevc.
The four final-state particles are required to form a good-quality vertex, where the resulting \Lb  candidate   is 
consistent with originating from a PV.
The vertex is also required to be significantly displaced from this PV. In order to reject the background from \jpsippi and \mbox{$\Lb\to\psi(2S)\proton\pim$} decays, the regions $8.0<\qsq<11.0$\gevgevcccc and \mbox{$12.5<\qsq<15.0$\gevgevcccc} are excluded from the signal search, where \qsq refers to the invariant mass squared of the two muons. In addition, contributions from \mbox{$\Lb\to\PLambda^{0}(\to\proton\pim)\mumu$} decays are removed by requiring $m_{p\pi^-}>1.12\gevcc$. 

Several fully reconstructed decays with at least one misidentified particle
can form backgrounds that peak in the distribution of the $\proton\pim\mumu$ mass, $m_{\proton\pim\mumu}$. Specific vetoes
are used to reject such backgrounds. The vetoes require that if the invariant mass of the candidate
is consistent with a particular hypothesis, then a more restrictive PID requirement is applied. 
For example, if the proton candidate is assigned the kaon mass and falls within the mass range $5246<m_{K^{+}\pim\mumu}<5330\mevcc$, the PID cut is significantly tightened to reduce $K\to \proton$ misidentification from $\Bz\to K^{+} \pi^{-} \mumu$ decays. 
Other possible sources of specific backgrounds are the decays $\pkmumu$, $\Bs\to \pi^{+}\pi^{-} \mumu$, \mbox{$\Bs\to K^{+} K^{-} \mumu$} and crossfeed from the normalisation channel \jpsippi. After the vetoes have been
applied, the only significant residual background contribution for the signal (normalisation) channel comes from the decay $\pkmumu$ (\mbox{$\Lb\to\jpsi(\to\mumu)\proton\kaon^{-}$}). This contamination is treated as a systematic uncertainty in the signal channel and is considered explicitly when extracting the yield of the normalisation channel. Partially reconstructed \mbox{$\Lb\to(\Lc\to\proton\pim\mup\nu)\mun\nu$} decays, which contribute at 
$m_{\proton\pim\mumu}$ below the nominal $\Lb$ mass, are also explicitly considered when determining the signal yield. 

A boosted decision tree (BDT)~\cite{Breiman}, with the AdaBoost 
algorithm~\cite{AdaBoost} and a five-fold cross-validation method~\cite{Stone:1974}, 
is used to reduce combinatorial background. The BDT is trained and optimised on data. 
Candidates with $m_{\proton\pim\mumu} > 6000\mevcc$ are used as a sample representative of the background, and  $\Lb\to\jpsi(\to\mumu)\proton\kaon^{-}$ candidates selected from the data are used as a proxy for the signal sample. The BDT uses kinematic, geometric 
and PID variables associated with the proton to discriminate between the signal and background candidates. The two most discriminating input variables are the 
vertex quality of the \Lb candidate and its consistency with originating from a PV. 
In order to reject background containing additional tracks in close proximity to the \Lb vertex, an isolation parameter~\cite{LHCb-PAPER-2015-025} is also used as an input variable.
As the presence of a proton from a displaced vertex is a distinctive 
signature, PID information on the proton candidate is used in the BDT in order to improve the
rejection of background. Other, less discriminating variables used in the BDT include
 the minimum impact parameter with respect to any PV and the momenta of the final-state particles.
The requirement on the BDT response is optimised by maximising the figure of merit~\cite{Punzi:2003bu} defined as
\begin{displaymath}
 P = \frac{\varepsilon_{\mathrm{sel}}}{3/2  + \sqrt{B}},
\end{displaymath}
where $\varepsilon_{\mathrm{sel}}$ is the selection efficiency for the signal and $B$ is the background expected within 40\mevcc of the \Lb mass. After candidates have been reconstructed and the above selection criteria have been applied, the requirement on the BDT output retains 65\% of signal events and rejects 99\% of the background.

%% file: normalisation.tex
% $Id: introduction.tex 87303 2016-02-08 13:44:29Z lafferty $

\section{Normalisation}
\label{sec:Normalisation}

The branching fraction of \ppimumu can be determined from
\begin{equation}
  \begin{split}
    \mathcal{B}(\ppimumu) &= \mathcal{B}(\ppijpsi)\mathcal{B}(\jpsi\to\mumu)\times\\
    &\phantom{=} \frac{N(\ppimumu)}{N(\jpsippi)}\frac{\varepsilon(\jpsippi)}{\varepsilon(\ppimumu)},\\
\end{split}
  \label{eq:bf}
\end{equation}
where $N(X)$ is the yield of the final state $X$ and $\varepsilon(X)$ is the efficiency to 
select that final state. The efficiencies are obtained from simulated events and specific control samples in the data. 
Since the normalisation channel \jpsippi has the same final state and 
similar kinematics as the signal decay, many systematic uncertainties cancel 
in the efficiency ratio. 

Control channels selected from the data are used to account  for several effects that are mismodelled in the simulation.
For example, the PID efficiencies are obtained from data samples with decays where the final-state particles can be identified by kinematic constraints alone \cite{LHCb-DP-2012-003}. Further corrections are derived by comparing the data and simulation distributions of the $\Lb$ momentum, transverse momentum, decay time and the track multiplicity for the normalisation channel. The relative efficiency of the BDT is calculated using both \jpsipk and \pkmumu candidates selected from the data; the resulting efficiencies are consistent with each other. The most important difference in the efficiency between the signal and normalisation modes is due to the \qsq selection for the signal decay, which removes 30\% of the signal candidates. For the full selection, including the dimuon mass vetoes, the total relative efficiency is found to be ${\varepsilon(\ppimumu)}/{\varepsilon(\jpsippi)}=\tw$.

\begin{figure}[tb]
  \centering
  \includegraphics[width=0.5\linewidth]{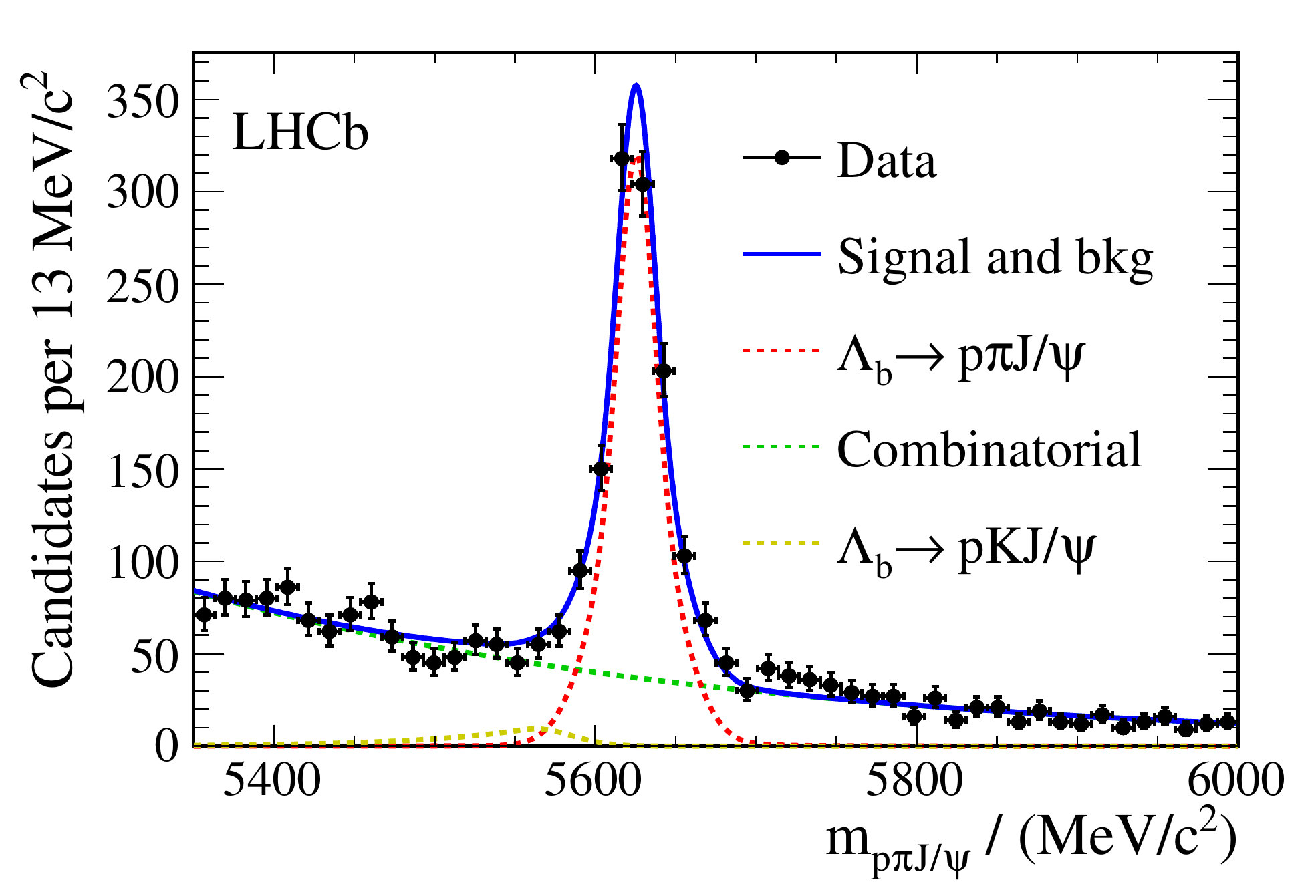}
\vspace*{-0.5cm}
  \caption{Mass distribution of \jpsippi candidates compared to the result of the fit. The fit parameterisation is described in the text.}
  \label{fig:jpsippi}
\end{figure}

 For the normalisation channel, candidates are required to have a dimuon mass within 
 60\mevcc of the known \jpsi\xspace mass. The yield of the normalisation channel 
 is obtained by performing an extended unbinned maximum likelihood
 fit to the \jpsippi mass distribution, as shown in Fig.~\ref{fig:jpsippi}. The shape of the \jpsippi mass distribution is described by the sum of 
 two Gaussian functions with power law tails and a shared mean, where the Gaussian parameters are allowed
  to vary in the fit and the tail parameters are obtained from the simulation.  Combinatorial background
  is parameterised with an exponential function with a decay constant that is allowed to vary in the fit.
  Finally, there is a small contribution from the decay \jpsipK, the shape of which is determined
  from the simulation and included in the fit to the data. In total,  $\Nyield$ \jpsippi candidates are observed. This yield is significantly lower than
 in Refs~\cite{LHCB-PAPER-2014-020,LHCb-PAPER-2016-015}, owing to the tighter selection employed to search for the \ppimumu decay. 
  
%% file: results.tex
% $Id: introduction.tex 87303 2016-02-08 13:44:29Z lafferty $

\section{Results}
\label{sec:Results}

\begin{figure}[!htb]
  \centering
  \includegraphics[width=0.5\linewidth]{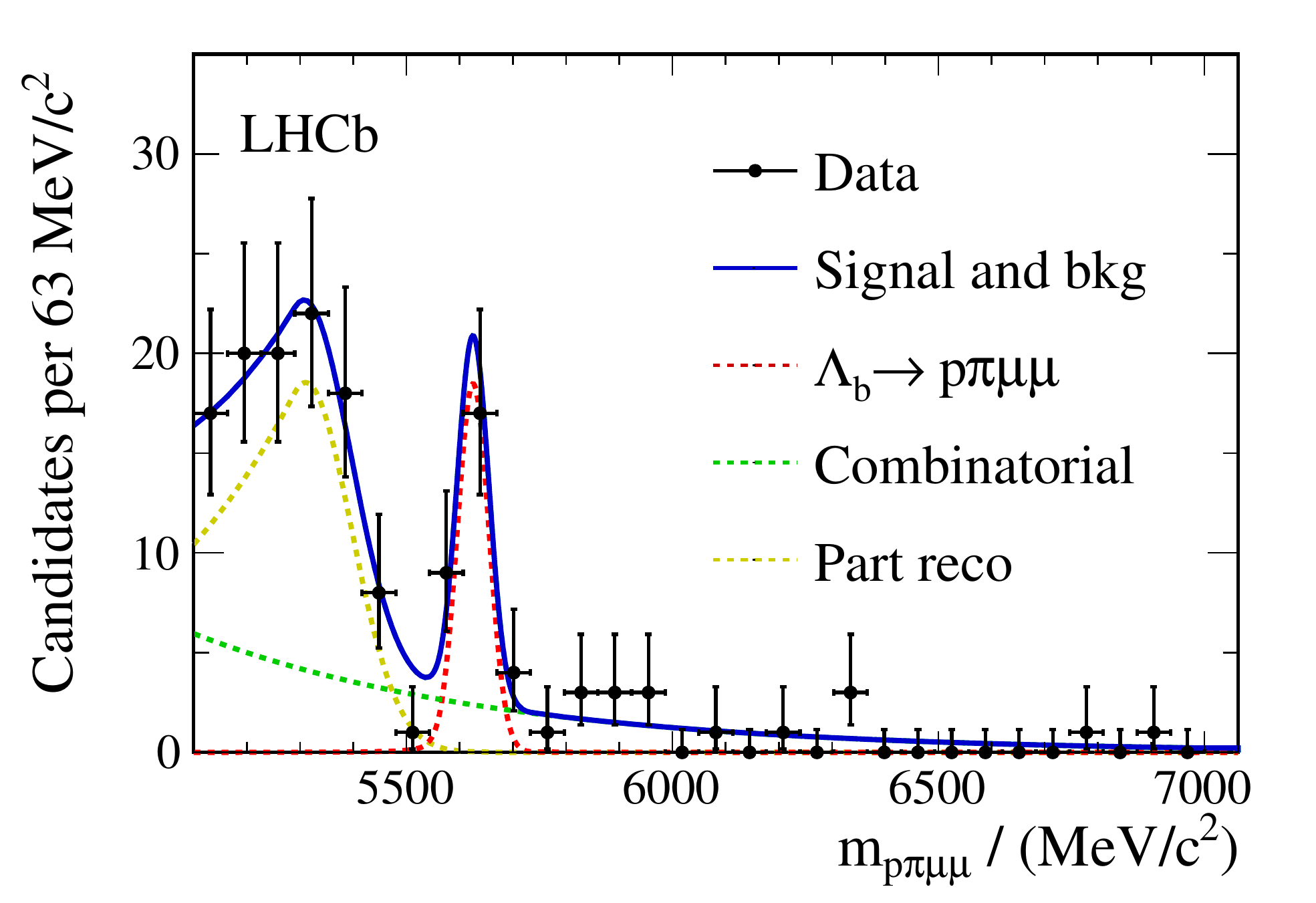}
  \vspace*{-0.5cm}
  \caption{Mass distribution of \ppimumu candidates compared to the result of the fit. The fit parameterisation is described in the text.}
  \label{fig:ppimumu}
\end{figure}

The fit to the invariant mass distribution of \ppimumu candidates, excluding the \jpsi and \psitwos regions, is shown in Fig.~\ref{fig:ppimumu}. The signal shape
is determined from the fit to the normalisation decay in data, with corrections for the differences between the signal and normalisation modes obtained from the simulation. 
The combinatorial background is parameterised as in the fit for the normalisation mode. The shape of the partially reconstructed background is obtained from a fit to the \pkmumu mass spectrum and the yield is allowed to vary in the fit to the \ppimumu mass distribution.

A signal contribution is clearly visible and Wilks' theorem \cite{wilks1938} gives a  significance of $\SIG$ standard deviations. The systematic uncertainties described in Sec.~\ref{sec:Systematics} are mainly associated with the normalisation. Only the systematic uncertainty arising from the shape assumed for the partially reconstructed background has any appreciable impact on the significance. Releasing the constraints on the relevant parameters, the significance \emph{increases} to 5.7 standard deviations. Pseudoexperiments indicate that, on-average, the significance would be expected to decrease by 0.3 standard deviations. Given the statistical variation, the observed increase is perfectly compatible with the expectation. 
This analysis therefore constitutes the first observation of the decay \ppimumu.
The number of signal candidates is found to be $\Syield$, 
which is converted to relative and absolute branching fractions of
\begin{align*}
\nonumber \frac{\mathcal{B}(\ppimumu)}{\mathcal{B}({\decay{\Lb}{ \jpsi(\to\mumu)\proton\pim}})} &=  0.044\pm0.012\pm0.007 \\
\nonumber \shortintertext{and}
\nonumber \mathcal{B}(\ppimumu) &= \BFV
\end{align*}
using Eq.~\ref{eq:bf}. In both cases, the first uncertainty given is statistical and the second is the systematic uncertainty, which is discussed in the next section. 
The third uncertainty on \mbox{$\mathcal{B}(\ppimumu)$} arises from the 
limited knowledge of the \jpsippi\cite{LHCB-PAPER-2014-020,LHCB-PAPER-2015-032} and \jpsi\to\mumu\cite{Olive:2016xmw} branching fractions.

%% file: systematics.tex
% $Id: introduction.tex 87303 2016-02-08 13:44:29Z lafferty $

\section{Systematic uncertainties}
\label{sec:Systematics}

The systematic uncertainties are summarised in Table~\ref{Tab:sys}. 
The total systematic uncertainty is $16.1\%$, which is comparable to but smaller than the statistical uncertainty. 

\newcommand\Tstrut{\rule{0pt}{2.6ex}}
\newcommand\TTstrut{\rule{0pt}{3.2ex}}
\newcommand\Bstrut{\rule[-1.2ex]{0pt}{0pt}}
\newcommand\BBstrut{\rule[-1.8ex]{0pt}{0pt}}

\begin{table}[ht]

  \centering
  \caption{Summary of the systematic uncertainties on the \ppimumu branching fraction.}
  \begin{tabular}{l c c c c}
    \hline
    Source & Uncertainty (\%) \\

    \hline
    Modelling of the $q^{2}$ distribution\Tstrut &  7.9  \\
        Modelling of the $\proton{}\pim$ mass distribution&  7.7  \\
    Effect of the partially reconstructed background fit shape &  6.9  \\
    Choice of BDT efficiency proxy&  5.6  \\
    Finite size of the simulated sample &  4.4  \\
    Statistical uncertainty on \jpsippi yield & 4.0\\
    Trigger efficiency &  3.4    \\
    Fit bias & 2.2 \\
    \pKmumu contamination &  1.6  \\
    Simulation corrections&  1.3  \\
    PID efficiency  &  1.0  \\
    \hline

    Total \Tstrut & 16.1 \\\hline

  \end{tabular}

  \label{Tab:sys}
\end{table}
The  largest systematic uncertainty originates from the decay model used to simulate the signal. 
There are two components to this uncertainty. The first originates from the unknown \qsq distribution
 for the signal decay.  As no model for the 
\ppimumu decay currently exists, the model for the decay $\Lb\to\PLambda^{0}(\to \proton{}\pim)\mumu$ from Ref.~\cite{Meinel} is used to derive the \qsq distribution. To assess the systematic uncertainty from this assumption, the decay \pkmumu is instead assumed to describe the signal \qsq distribution and the difference in relative efficiency is assigned as a systematic uncertainty. The \qsq distribution for the \pkmumu decay is obtained from data weighted using the \sPlot technique \cite{Pivk:2004ty}. An uncertainty of 7.9\% is found.
The second component of the systematic uncertainty due to the decay model is the distribution of the $\proton{}\pim$ invariant mass. In this case, the distribution in the simulation
 is weighted to match the data for the \ppijpsi decay and the efficiency is reevaluated. The difference of 7.7\% in relative efficiency between 
 these two cases is taken as a systematic uncertainty.
 
 Another important source of systematic uncertainty is related to the assumption
 that the partially reconstructed background for the signal has the same shape 
 as the partially reconstructed background in \pkmumu decays. The effect of this assumption is estimated by
  allowing the shape parameters for the partially reconstructed background component to vary in the fit, and then calculating the resulting bias in the 
  background estimation using pseudoexperiments. This results in a 6.9\% uncertainty on the signal yield. As noted above, this is the only systematic uncertainty that has an appreciable effect on the significance for the observation of the decay \ppimumu. 
  
Other, smaller uncertainties are assigned to the calculation of the efficiency: the calibration of the BDT efficiency using data (5.6\%);  the finite size of the simulation samples used (4.4\%) and possible mismodelling of the trigger (3.4\%). The statistical uncertainty on the \jpsippi yield gives rise to a systematic uncertainty of 4.0\%. 
Due to the low number of signal candidates, a small bias
  in the signal yield is observed. The size of this bias is calculated using pseudoexperiments and results in a 2.2\% systematic uncertainty. 
  No \pkmumu contribution is considered for the \ppimumu fit, due to the low expected yield. The \pkjpsi decay is used to assess the resulting systematic uncertainty,  which is 1.6\%. The corrections applied to the simulation give rise to a small systematic uncertainty (1.3\%), as does the calibration of the PID efficiency using data (1.0\%).

%% file: conclusions.tex
% $Id: introduction.tex 87303 2016-02-08 13:44:29Z lafferty $

\section{Conclusions}
\label{sec:Conclusions}

A search for the rare decay \ppimumu has been performed with proton-proton collision data collected with the LHCb experiment corresponding to 3\invfb of integrated luminosity. The search is made excluding the \jpsi~and \psitwos\to\mumu resonances. A signal is observed with a significance
of $\SIG$ standard deviations, which constitutes the first observation of a  
$\bquark\to\dquark$ transition in a baryonic decay. The relative and absolute branching fractions are measured to be

\begin{align*}
\nonumber
  \frac{\mathcal{B}(\ppimumu)}{\mathcal{B}( {\decay{\Lb}{ \jpsi(\to\mumu)\proton\pim}})} &=  0.044\pm0.012\pm0.007 \\
\nonumber \shortintertext{and}
  \nonumber
\mathcal{B}(\ppimumu) &= \BFV,
\end{align*}
where the first uncertainties are statistical and the second are systematic. The third uncertainty on $\mathcal{B}(\ppimumu)$ arises from the 
limited knowledge of the \mbox{\jpsippi}~\cite{LHCB-PAPER-2014-020} and \jpsi\to\mumu~\cite{Olive:2016xmw} branching fractions. 
With further advances in lattice QCD combined with a \pkmumu branching fraction measurement, 
this result will allow $|\Vtd|/|\Vts|$ to be measured, enabling a test of the Minimal Flavour Violation hypothesis.

%% file: acknowledgements.tex
\section*{Acknowledgements}
 
\noindent We express our gratitude to our colleagues in the CERN
accelerator departments for the excellent performance of the LHC. We
thank the technical and administrative staff at the LHCb
institutes. We acknowledge support from CERN and from the national
agencies: CAPES, CNPq, FAPERJ and FINEP (Brazil); MOST and NSFC (China);
CNRS/IN2P3 (France); BMBF, DFG and MPG (Germany); INFN (Italy); 
FOM and NWO (The Netherlands); MNiSW and NCN (Poland); MEN/IFA (Romania); 
MinES and FASO (Russia); MinECo (Spain); SNSF and SER (Switzerland); 
NASU (Ukraine); STFC (United Kingdom); NSF (USA).
We acknowledge the computing resources that are provided by CERN, IN2P3 (France), KIT and DESY (Germany), INFN (Italy), SURF (The Netherlands), PIC (Spain), GridPP (United Kingdom), RRCKI and Yandex LLC (Russia), CSCS (Switzerland), IFIN-HH (Romania), CBPF (Brazil), PL-GRID (Poland) and OSC (USA). We are indebted to the communities behind the multiple open 
source software packages on which we depend.
Individual groups or members have received support from AvH Foundation (Germany),
EPLANET, Marie Sk\l{}odowska-Curie Actions and ERC (European Union), 
Conseil G\'{e}n\'{e}ral de Haute-Savoie, Labex ENIGMASS and OCEVU, 
R\'{e}gion Auvergne (France), RFBR and Yandex LLC (Russia), GVA, XuntaGal and GENCAT (Spain), Herchel Smith Fund, The Royal Society, Royal Commission for the Exhibition of 1851 and the Leverhulme Trust (United Kingdom).

%% file: LHCb_Authorship_flat_27-Sep-2016.tex
\centerline{\large\bf LHCb collaboration}
\begin{flushleft}
\small
R.~Aaij$^{40}$,
B.~Adeva$^{39}$,
M.~Adinolfi$^{48}$,
Z.~Ajaltouni$^{5}$,
S.~Akar$^{59}$,
J.~Albrecht$^{10}$,
F.~Alessio$^{40}$,
M.~Alexander$^{53}$,
S.~Ali$^{43}$,
G.~Alkhazov$^{31}$
% ...

{\footnotesize \it
% ...
$ ^{5}$Clermont Universit{\'e}, Universit{\'e} Blaise Pascal, CNRS/IN2P3, LPC, Clermont-Ferrand, France\\
% ...
$ ^{10}$Fakult{\"a}t Physik, Technische Universit{\"a}t Dortmund, Dortmund, Germany\\
% ...
$ ^{31}$Petersburg Nuclear Physics Institute (PNPI), Gatchina, Russia\\
% ...
$ ^{39}$Universidad de Santiago de Compostela, Santiago de Compostela, Spain\\
$ ^{40}$European Organization for Nuclear Research (CERN), Geneva, Switzerland\\
% ...
$ ^{43}$Nikhef National Institute for Subatomic Physics, Amsterdam, The Netherlands\\
% ...
$ ^{48}$H.H. Wills Physics Laboratory, University of Bristol, Bristol, United Kingdom\\
% ...
$ ^{53}$School of Physics and Astronomy, University of Glasgow, Glasgow, United Kingdom\\
% ...
$ ^{59}$University of Cincinnati, Cincinnati, OH, United States\\
% ...
}
\end{flushleft}